\documentclass{isprs}
\usepackage{subfigure}
\usepackage{setspace}
\usepackage{geometry}
\usepackage{epstopdf}
\usepackage[labelsep=period]{caption}
\usepackage[british]{babel} 
\usepackage[hang]{footmisc}
\usepackage[colorlinks=false, pdfborder={0 0 0}]{hyperref}
\usepackage{fancyhdr}
\usepackage{amsmath}
\usepackage{tabularx}
\usepackage{booktabs}
\usepackage{lettrine}

\begin{document}
\title{
    \Large
    \textit{
        A Theoretical Framework for Environmental Similarity and Vessel\\Mobility as Coupled Predictors of Marine Invasive Species Pathways
    }
    \vspace{-.4cm}
}
\author{
    \large 
    Gabriel Spadon, Vaishnav Vaidheeswaran, Claudio DiBacco
    \vspace{-.2cm}
}
\address{
    Faculty of Computer Science, Dalhousie University, Halifax -- NS, Canada\\
    Fisheries and Oceans Canada, Bedford Institute of Oceanography, Dartmouth -- NS, Canada\\
    \{spadon, vaishnav\}@dal.ca, claudio.dibacco@dfo-mpo.gc.ca\\
    \vspace{-1.2cm}
}
\maketitle
\sloppy
\lettrine[lines=2, lhang=0.05, loversize=0.01]{M}{}arine invasive species transported via global shipping pose critical threats to biodiversity and cause significant economic damage, with documented losses reaching at least US\$23 billion in 2020~\cite{cuthbert2021global}. Shipping accounts for 80\% of world trade, with ballast water discharge and hull biofouling as primary vectors of spread~\cite{costello2022assessing}. Traditional risk assessment methods rely on detailed ballast water discharge records and shipping traffic data that are often unavailable or inaccessible, which limits comprehensive global risk mapping. An alternative approach uses the ecological principle of \textit{environmental matching}, which states that species are more likely to thrive in regions climatically similar to their native habitats. Environmental similarity, particularly sea surface temperature, salinity, and seasonal patterns, is a robust predictor of invasion success independent of shipping volume data~\cite{tzeng2024quantifying}. Under climate change, warming oceans are reshaping these environmental vulnerabilities, particularly in Arctic and high-latitude regions where retreating sea ice and rising temperatures enable species to prevail~\cite{chan2019climate}.

Advances in GeoAI\footnote{Geospatial Artificial Intelligence} enable the integration of heterogeneous big data with machine learning for spatial modeling and decision making. Massive streams from ship tracking data, satellite imagery, hydrophone arrays, and reanalysis products can be time-aligned, georeferenced, and encoded as tensors or graphs, enabling representation learning that yields embeddings of ports, routes, and seasons that preserve environmental and behavioral structure. On these embeddings, scalable clustering and metric learning expose climate analogues~\cite{kurihana2024identifying}, while probabilistic classifiers produce calibrated risk estimates with uncertainty that can be propagated through downstream analyses~\cite{koldasbayeva2024challenges}. The resulting models support policy by ranking source-to-sink pathways~\cite{bradie2021decision}, evaluating interventions such as routing or ballast protocols~\cite{nfongmo2024estimation}, and optimizing monitoring under budget and coverage constraints, which transforms data volume into operational guidance.

Understanding and quantifying these invasion pathways requires determining how environmental similarity and maritime mobility jointly influence the probability of species transfer among all publicly registered anchorage sites (hereinafter ``ports"). This paper presents a theoretical formulation\footnote{Full implementation details will be reported in an extended publication.} of the problem and advances a globally based yet regionalized framework for policy and management applications in invasive species risk quantification. We model environmental similarity across global ports, where each location $p_i$ is represented by a standardized feature vector $\mathbf{x}_i=[x_{i1},x_{i2},\ldots,x_{id}]$ describing local marine climate. Let $x_{i,v,t}$ denote the value of variable $v$ at port $i$ and month $t\in\{0,\ldots,11\}$. Features include mean conditions $\mu_{i,v}$, seasonal cycles $s_{i,v}(t)$, temporal variability $\operatorname{Var}[x_{i,v,\cdot}]$, and extremes $\max_t x_{i,v,t}$ and $\min_t x_{i,v,t}$. Ports are compared through a distance $d(\mathbf{x}_i,\mathbf{x}_j)$, such as Euclidean.

For each port $p_i$, we phase align monthly records on $t'=(t+\delta_i)\bmod 12$, where $\delta_i$ shifts the series to a common seasonal anchor (for opposite hemispheres, $\delta_i\in\{0,6\}$). From the aligned series, we compute a feature vector $\tilde{\mathbf{x}}_i$ that summarizes means, seasonal amplitudes and phases, variability, and extremes. HDBSCAN\footnote{\url{https://hdbscan.readthedocs.io/en/latest/index.html}} is applied to the aligned representations $\{\tilde{\mathbf{x}}_i\}_{i=1}^{N}$ using a metric $d(\tilde{\mathbf{x}}_i,\tilde{\mathbf{x}}_j)$ that captures environmental dissimilarity. The algorithm yields clusters $\mathcal{C}_k=\{\,p_i\mid \ell(p_i)=k\,\}$ that group ports with comparable climatic traits while recognizing transitional and outlier environments where density is insufficient for stable assignment. Climate matching uses the aligned features, $S(p_i,p_j)=\frac{1}{1+d(\tilde{\mathbf{x}}_i,\tilde{\mathbf{x}}_j)}\in(0,1]$, which evaluates similarity on comparable phenological phases.

Maritime mobility provides the transport mechanism linking environmental analogues, and Automatic Identification System (AIS) messages provide the observational backbone for reconstructing vessel movements. Each message is a time-stamped tuple $m=(\mathrm{MMSI},t,\phi,\lambda,\mathrm{sog},\mathrm{cog})$ recording ship identity, time, position, speed over ground, and course over ground, which we aggregate into voyages and port calls. We represent the transport system as a directed graph $G=(V,E)$, where nodes $V=\{p_i\}$ are ports and an edge $e_{ij}\in E$ exists when one or more vessels travel from $p_i$ to $p_j$ within the observation window, with weight $w_{ij}$ proportional to voyage frequency or tonnage. Nodes inherit cluster labels $\ell(p_i)\in\{1,\ldots,K\}$ from the environmental stage. A voyage with path $\pi=(i_0,\ldots,i_H)$ from donor port $p_{i_0}$ to recipient port $p_{i_H}$ carries a conditional risk summarized by a shipment score $\rho(\pi)$ that depends on environmental similarity $S(p_{i_{h-1}},p_{i_h})$ along the path and voyage specific factors. Risk transmitted along a route of $H$ hops decays as $\gamma^{\,h-1}$ with $0<\gamma\le 1$, so multi-hop pathways dilute per-voyage risk yet can accumulate when exposed repeatedly.

Using historical global AIS data, we train a temporal link-prediction model to project how vessel traffic may reconfigure as climate-affected regions shift. Let $G_t=(V,E_t,W_t)$ denote the port mobility network at time $t$, with directed edges $E_t$ and edge weights $w_{ij,t}$. For each ordered pair $(i,j)$ we form edge features $z_{ij,t}$ that combine mobility history (lags of $w_{ij,t}$ and route recency), environmental correspondence ($S(p_i,p_j)$ and its change $\Delta S_{ij,t}$ under a scenario), node attributes (cluster labels $\ell(p_i),\ell(p_j)$, port capacity), and exogenous covariates $x^{\mathrm{exo}}_{i,t},x^{\mathrm{exo}}_{j,t}$ including econometric, demographic, and policy indicators. The target is a future link outcome $y_{ij,t+\Delta}$, defined either as a binary variable that equals $1$ if $w_{ij,t+\Delta}>\tau$ and $0$ otherwise, or as a scaled intensity. We fit neural temporal models $f_\theta$ that map histories $z_{ij,1:t}$ to $\hat y_{ij,t+\Delta}$ and benchmark against baselines (e.g., logistic regression, gradient boosting, and random forests) and ensembles $\hat y^{\mathrm{ens}}_{ij}=\sum_m \alpha_m \hat y^{(m)}_{ij}$.

To combine environmental similarity with predicted mobility, let $\hat Y_{t+\Delta}=[\hat y_{ij,t+\Delta}]$ be the link prediction matrix. Map similarity into an edge level risk weight via an environmental kernel $\mathcal{K}=[\kappa_{ij}]$ with $\kappa_{ij}=\phi\!\big(S(p_i,p_j)\big)\,\bigl(1+\beta\,\delta_{\ell(p_i),\ell(p_j)}\bigr)$, where $\phi$ is monotone (e.g., $\phi(u)=u^\eta$ with $\eta>0$), $\beta\ge 0$ modulates within class reinforcement, and $\delta$ is the Kronecker delta. The traffic conditioned risk adjacency is $A=\hat Y_{t+\Delta}\odot\mathcal{K}$. One hop inbound exposure at recipient port $p_r$ is $E^{(1)}_{r,t+\Delta}=\big(A^\top\mathbf{1}\big)_r$. To capture multi hop pathways we sum over walks of length $h=1,\ldots,H$ with geometric decay, $E_{r,t+\Delta}=\sum_{h=1}^{H}\gamma^{\,h-1}\,\big((A^{h})^\top\mathbf{1}\big)_r$, $0<\gamma\le 1$, which discounts distant routes while allowing cumulative exposure. For an individual vessel with observed path $\pi=(i_0,\ldots,i_H=r)$, shipment level risk complements survival along edges, $\rho(\pi)=1-\prod_{h=1}^{H}\Bigl(1-\gamma^{\,h-1}\,\kappa_{i_{h-1}i_h}\Bigr)$, optionally scaled by voyage factors such as residence time or ballast handling.

To illustrate the framework in practice, consider a hypothetical example for Nova Scotia as a recipient region. After phase alignment, Halifax Harbour \(p_{\mathrm{HFX}}\), Sydney \(p_{\mathrm{SYD}}\), and the Strait of Canso \(p_{\mathrm{CNS}}\) fall into a cold-temperate cluster that also contains northern European ports such as Rotterdam \(p_{\mathrm{RTM}}\) and Gothenburg \(p_{\mathrm{GOT}}\). The aligned features yield high pairwise similarity, for example \(S(p_{\mathrm{RTM}},p_{\mathrm{HFX}})\approx 0.80\) and \(S(p_{\mathrm{GOT}},p_{\mathrm{SYD}})\approx 0.77\), while similarity to subtropical donors remains low. AIS trajectories over twelve months reconstruct regular container service loops from \(p_{\mathrm{RTM}}\) and feeder routes via St.\ John's \(p_{\mathrm{SJN}}\) into Halifax and Sydney, producing edges \(i\to j\) with weights \(w_{ij}\) proportional to voyage counts. A temporal link model fitted on prior years' projects \(\hat{y}_{ij,t+\Delta}\) that increases transatlantic frequency in spring when thermal ranges align, and the environmental kernel \(\kappa_{ij}\) upweights within-cluster pairs. The resulting adjacency \(A=\hat{Y}\odot\mathcal{K}\) yields one hop exposure \(E^{(1)}_{t+\Delta}\) that peaks at Halifax in April to June, while multi hop exposure with \(\gamma=0.6\) highlights Sydney in September due to coastal redistribution from Halifax through Canso. Shipment level scores \(\rho(\pi)\) then flag individual voyages with long residence times at berths in Bedford Basin or Sydney Harbour combined with high \(\kappa_{ij}\). The decision output is a ranked list of vessel-port-month triplets for targeted inspections, ballast management, and hull surveys, together with evaluations that quantify how routing adjustments or ballast exchange rules reduce \(E\) by a chosen margin with uncertainty estimates.

Accordingly, the whole system emerges as a fusion of complementary data streams and interacting models. Environmental similarity networks derived from reanalysis and satellite products define the ecological background, while AIS-based mobility graphs capture the dynamics of transmission. Temporal predictors, probabilistic classifiers, and clustering methods each provide partial evidence that reflects distinct mechanisms of invasion risk. Through model fusion, these sources are integrated into a single inference pipeline that harmonizes information across spatial and temporal resolutions. The framework supports localized predictions of exposure and transport potential at both the port and vessel levels, with every ship and destination receiving an individualized risk estimate grounded in its climatic, operational, and network context. In doing so, the system transforms fragmented global observations into a coherent, site-specific intelligence framework that can guide surveillance priorities, adaptive routing, and policy interventions.

Although the framework shows promise, there are significant gaps in how AI interacts with ocean dynamics. Many models rely on climatological averages, overlooking important features such as fronts and eddies. Data integration is challenging due to mismatches in spatial and temporal scales across sources. We need methods to align these data streams, quantify uncertainties, and integrate findings into risk estimates. Additionally, limited interpretability hinders the traceability of results and policy implications. Addressing these gaps requires ocean-aware AI that adapts to evolving flow fields, combines ecological processes with transport, and integrates physical and ecological knowledge into a unified decision system.

In summary, this work presents a theoretical framework for quantifying marine invasion risk by jointly modeling environmental similarity and maritime mobility within a unified problem. By integrating satellite observations, reanalysis fields, and global vessel trajectories, it learns predictive representations that reveal pathways of biological transfer under a changing climate. The methodology scales from global networks to local ports and translates coupled oceanic and anthropogenic processes into localized, decision-ready intelligence for monitoring, routing, and management. Beyond invasive species, it outlines a broader paradigm for ocean analytics in which data fusion, model interoperability, and adaptive learning enable real-time environmental governance in dynamic marine systems.

\renewcommand{\refname}{\centering\MakeUppercase{REFERENCES}}
{
	\begin{spacing}{1.1}
		\normalsize
		\bibliography{ISPRSguidelines_authors}
	\end{spacing}
}

\section*{Acknowledgements}
This work was partially supported by the Natural Sciences and Engineering Research Council of Canada (NSERC) and the Brazilian National Council for Scientific and Technological Development (CNPq), under process 444325/2024-7.

\end{document}